\documentclass[twocolumn,showpacs]{revtex4} %
\usepackage{epsfig,amsmath,amssymb,graphicx,verbatim}
\usepackage{natbib}
\usepackage[english]{babel} 

\begin{document}

\vspace{0mm}
\title{ THERMODYNAMICS OF THE FERMI GAS IN A NANOTUBE} %

\author{Yu.M. Poluektov}
\email{yuripoluektov@kipt.kharkov.ua} %
\affiliation{National Science Center ``Kharkov Institute of Physics
and Technology'', Akhiezer Institute for Theoretical Physics, 61108
Kharkov, Ukraine}
\author{A.A. Soroka} %
\affiliation{National Science Center ``Kharkov Institute of Physics
and Technology'', Akhiezer Institute for Theoretical Physics, 61108
Kharkov, Ukraine}

\begin{abstract}
For the ideal Fermi gas that fills the space inside a cylindrical tube, %
there are calculated the thermodynamic characteristics %
in general form for arbitrary temperatures, namely: %
the thermodynamic potential, energy, entropy, equations of state,
heat capacities and compressibilities. All these quantities are
expressed through the introduced standard functions and their
derivatives. The radius of the tube is considered as an additional
thermodynamic variable. It is shown that at low temperatures in the
quasi-one-dimensional case the temperature dependencies of the
entropy and heat capacities remain linear. The dependencies of the
entropy and heat capacities on the chemical potential have sharp
maximums at the points where the filling of a new discrete level
begins. The character of dependencies of thermodynamic quantities on
the tube radius proves to be qualitatively different in the cases of
fixed linear and fixed total density. At the fixed linear density
these dependencies are monotonous and at the fixed total density they have
an oscillating character. \newline
\end{abstract}
\maketitle

\section{Introduction} 
The model of the ideal Fermi gas is the basis for understanding the
properties of electron and other many-fermion systems. In many cases
it is also possible to describe with reasonable accuracy the
behavior of systems of interacting Fermi particles within the
approximation of an ideal gas of quasiparticles whose dispersion law
differs from the dispersion law of free particles. It is essential
that thermodynamic characteristics of the ideal Fermi gas at
arbitrary temperatures in the volume case can be expressed through
the special Fermi functions, and thus it is possible to obtain and
verify all relations of the phenomenological thermodynamics on the
basis of the quantum microscopic model.

In recent time, much attention has been paid to investigation of
low-dimensional systems, in particular of properties of the Fermi gas %
in nanotubes and quantum wells, because apart from purely
scientific interest the study of such objects is rather promising
for the solid-state electronics \cite{Ando,Komnik,DNG,Vagner,Freik,SP}. %
Thermodynamics relations for the Fermi gas in the confined geometry
have been studied much less than in the volume case \cite{LL} and
require further investigation. Thermodynamic properties of the Fermi
gas at arbitrary temperatures in a quasi-two-dimensional quantum
well have been studied in detail by the authors in work \cite{PS}. A
detailed understanding of the properties of such systems must serve
as a basis for the study of low-dimensional systems of interacting
particles \cite{Òîmonaga,Luttinger,Deshpande}.

It is considered that strongly correlated Fermi systems, to
which also low-dimensional systems of Fermi particles are attributed, %
in many respects essentially differ from the usual Fermi liquid
systems \cite{Landau,PN}. At that the properties of
quasi-one-dimensional or quasi-two-dimensional systems are often
compared with the theory of bulk Fermi liquid \cite{Landau,PN}. %
However, as seen even on the examples of the quasi-two-dimensional
system of noninteracting particles which was considered by the
authors in work \cite{PS} and the quasi-one-dimensional system
considered in the present work, the properties of low-dimensional
systems can substantially differ, especially at low temperatures,
from the properties of the bulk system owing to the quantum size
effect. For this reason, the theory of Fermi liquid itself in
conditions of the confined geometry must, generally speaking, be
formulated differently than in the volume case.

The consideration of low-dimensional models of interacting Fermi
particles leads to a conclusion about unique properties of such
systems \cite{Òîmonaga,Luttinger,Deshpande}. It should be kept in
mind, however, that real systems are always three-dimensional and
their low dimensionality manifests itself only in the boundedness of
motion of particles in one, two or three coordinates.

In considering statistical properties of the three-dimensional
many-particle systems one usually passes in results to the
thermodynamic limit, setting in final formulas the volume and number
of particles to infinity at the fixed density. It is of general
theoretical interest to study the statistical properties of
many-particle systems in which one or two dimensions remain fixed,
and the thermodynamic limiting transition is carried out only over
remaining coordinates. In this case the coordinates, over which the
thermodynamic limiting transition is not performed, should be
considered as additional thermodynamic variables. The model of the
ideal Fermi gas allows to build the thermodynamics of
low-dimensional systems based on the statistical treatment.

In the present work, the thermodynamics of the ideal Fermi gas that
fills the space inside a cylindrical tube is studied. There are
calculated its thermodynamic characteristics in general form for
arbitrary temperatures, namely: the thermodynamic potential, energy,
entropy, equations of state, heat capacities and compressibilities.
All these quantities are expressed through the standard functions
introduced in the work and their derivatives. It is suggested the
representation of thermodynamic quantities in the dimensionless
``reduced'' form, which is convenient owing to the fact that it does
not contain geometric dimensions of the system. The case of low
temperatures and tubes of small radius is studied more in detail. It
is shown that at low temperatures in the quasi-one-dimensional case
under consideration the temperature dependencies of the entropy and
heat capacities remain linear, the same as in the
quasi-two-dimensional \cite{PS} and the volume cases \cite{Landau,PN},
except the specific points where the filling of a new discrete level
begins. Moreover, the dependencies of the entropy and heat
capacities on the chemical potential have sharp maximums at these
points. The behavior of thermodynamic quantities with the tube
radius proves to be qualitatively different, depending on whether
the linear or the total density is fixed. At the fixed linear density
these dependencies prove to be monotonous and at the fixed total
density they have an oscillating character.

\vspace{-0mm}
\section{Thermodynamics of the Fermi gas in a cylindrical tube} 
The model of the ideal Fermi gas is the basis for studying the bulk
properties of Fermi systems of particles of different nature. %
In the two-dimensional case an analogous role is played by the ideal
Fermi gas contained between two parallel planes, whose
thermodynamics was considered in detail in \cite{PS}. %
In this work the thermodynamics properties of the ideal Fermi gas
enclosed in a cylindrical tube of radius $R$ and length $L$ are
studied. The length of the tube is assumed to be macroscopic, and no
restrictions are imposed on the radius of the tube. %
The main attention is paid to studying the properties of tubes of
small radius, which corresponds to transition to the
quasi-one-dimensional case, at low temperatures. It is everywhere
assumed that the spin of the Fermi particle is equal to 1/2. %
It is also assumed that the potential barrier on the surface of the
tube is infinite, so that the wave function of a particle turns into
zero at the boundary. The solutions of the Schr\"{o}dinger equation
in this case have the form
\begin{equation} \label{01}
\begin{array}{l}
\displaystyle{%
  \varphi_{k,n,\nu}(r,\alpha,z)=\frac{1}{J_n'(\rho_{n\nu})R\sqrt{\pi L}}\,e^{ikz}e^{\pm in\alpha}J_n\left(\rho_{n\nu}\frac{r}{R}\right), %
}%
\end{array}
\end{equation}
where $\rho_{n\nu}$ are the zeros of the Bessel function of the
order $n$, $J_n(\rho_{n\nu})=0$, $n=0,1,\ldots$, $\nu=1,2,\ldots$.
The energy of a particle:
\begin{equation} \label{02}
\begin{array}{l}
\displaystyle{%
  \varepsilon_{k,n,\nu}=\frac{\hbar^2k^2}{2m} + \rho_{n\nu}^2\frac{\hbar^2}{2mR^2}, %
}%
\end{array}
\end{equation}
and the distribution function has the form
\begin{equation} \label{03}
\begin{array}{ll}
\displaystyle{%
  f_{k,n,\nu}=\big[\exp\beta(\varepsilon_{k,n,\nu}-\mu) +1\big]^{-1},  %
}
\end{array}
\end{equation}
where $\beta=1/T$ is the inverse temperature. %
Pay attention that the discrete energy levels with $n=0$ are
nondegenerate, and the levels with $n\geq 1$ are twice degenerate in
the projection of angular momentum on the $z$ axis.

Thermodynamic functions of the Fermi gas both in the volume case and
in the case of lower dimensionality \cite{PS} can be expressed
through the Fermi function defined by the formula
\begin{equation} \label{04}
\begin{array}{l}
\displaystyle{%
  \Phi_s(t)=\frac{1}{\Gamma(s)}\int_0^{\!\infty} \frac{z^{s-1}\,dz}{e^{z-t}+1}, %
}%
\end{array}
\end{equation}
where $s$ is an integer or half-integer positive number, $\Gamma(s)$
is the gamma function. For calculation of the bulk properties of the
Fermi gas it is sufficient to know the functions (\ref{04}) with
half-integer indices $s=1/2,3/2,5/2$, and for the
quasi-two-dimensional case -- with integer indices $s=1,2$. For the
quasi-one-dimensional case under consideration in this work the
functions with half-integer indices $s=1/2,3/2$ are required as
well.

It is convenient to introduce the dimensionless temperature
$\tau\equiv T/\varepsilon_R$ and the dimensionless chemical
potential $\eta\equiv \mu/\varepsilon_R$, with normalization on the energy
\begin{equation} \label{05}
\begin{array}{l}
\displaystyle{%
  \varepsilon_R=\frac{\hbar^2}{2mR^2}, %
}%
\end{array}
\end{equation}
and to define the functions of these two dimensionless variables by
the following relation
\begin{equation} \label{06}
\begin{array}{ll}
\displaystyle{%
  \Psi_s(\tau,\eta)\equiv\sum_{n=0}^\infty\sum_{\nu=1}^\infty g_n\Phi_s(t_{n\nu}) = %
}\vspace{2mm}\\ %
\displaystyle{\hspace{13mm}%
  =\sum_{\nu=1}^\infty \Phi_s(t_{0\nu}) + 2\sum_{n=1}^\infty\sum_{\nu=1}^\infty \Phi_s(t_{n\nu}). %
}
\end{array}
\end{equation}
Here the parameter
$t_{n\nu}\equiv\beta(\mu-\rho_{n\nu}^2\varepsilon_R)\equiv\tau^{-1}(\eta-\rho_{n\nu}^2)$, %
and the factor $g_n\equiv 2-\delta_{n0}$ accounts for the degeneracy
order of levels. It equals unity for the levels with $n=0$ and two
for the rest of the levels.

After integration over momenta the thermodynamic potential $\Omega$,
number of particles $N$, energy $E$ and entropy $S$ are expressed
through the functions (\ref{06}) 
\begin{equation} \label{07}
\begin{array}{ll}
\displaystyle{%
  \Omega=-\frac{2LT}{\Lambda}\Psi_{3/2}(\tau,\eta), %
}
\end{array}
\end{equation}\vspace{-3mm}
\begin{equation} \label{08}
\begin{array}{ll}
\displaystyle{%
  N=\frac{2L}{\Lambda}\Psi_{1/2}(\tau,\eta), %
}
\end{array}
\end{equation}\vspace{-3mm}
\begin{equation} \label{09}
\begin{array}{ll}
\displaystyle{%
  E=\frac{LT}{\Lambda}\!\left[ \Psi_{3/2}(\tau,\eta) + 2\frac{\eta}{\tau}\Psi_{1/2}(\tau,\eta) + %
  2\tau\frac{\partial\Psi_{3/2}(\tau,\eta)}{\partial\tau} \right], %
}
\end{array}
\end{equation}\vspace{-3mm}
\begin{equation} \label{10}
\begin{array}{ll}
\displaystyle{%
  S=\frac{3L}{\Lambda}\!\left[ \Psi_{3/2}(\tau,\eta) + \frac{2}{3}\,\tau\frac{\partial\Psi_{3/2}(\tau,\eta)}{\partial\tau} \right], %
}
\end{array}
\end{equation}
where $\Lambda\equiv\left(2\pi\hbar^2\big/mT\right)^{\!1/2}$ is the
thermal de Broglie wavelength. The volume $n=N\big/\pi R^2L$ and the
linear $n_{\!L}=N\big/L$ densities of particles are obviously given as 
\begin{equation} \label{11}
\begin{array}{ll}
\displaystyle{%
  n=\frac{2}{\pi R^2\Lambda}\Psi_{1/2}(\tau,\eta), \qquad n_{\!L}=\frac{2}{\Lambda}\Psi_{1/2}(\tau,\eta). %
}
\end{array}
\end{equation}
The thermodynamic potential (\ref{07}) is a function of the
temperature, chemical potential, the length of the tube and its
radius: $\Omega=\Omega(T,\mu,L,R)$. In contrast to the volume case
when $\Omega$ is proportional to the volume $V$, in this case it is
proportional to the length $L$ and depends in a complicated manner
on the radius $R$. This circumstance is conditioned by the evident
anisotropy of the system under consideration, since here the motions
along the tube and in the direction of radius are qualitatively
different. In statistical mechanics it is customary to pass in the
final formulas to the thermodynamic limit: $V\rightarrow\infty$,
$N\rightarrow\infty$ at $n=N/V=\rm{const}$. In the present case it
is more accurate to write down the thermodynamic limit somewhat
differently, namely
\begin{equation} \label{12}
\begin{array}{ll}
\displaystyle{%
  L\rightarrow\infty,\, N\rightarrow\infty \textrm{\,\,\,\,at\,\,\,} n_{\!L}\equiv N/L=\rm{const}. %
}
\end{array}
\end{equation}
It is thereby stressed that the transition to infinite volume occurs
only owing to increasing the length of the tube,  with its radius
$R$ being fixed.

The differential of the thermodynamic potential (\ref{07}) has the
form
\begin{equation} \label{13}
\begin{array}{ll}
\displaystyle{%
  d\Omega= \frac{\Omega}{L}\,dL\,
  -\frac{4LT}{\Lambda}\!\left[ \frac{\eta}{\tau}\Psi_{1/2}+ \tau\frac{\partial\Psi_{3/2}}{\partial\tau} \right]\!\frac{dR}{R}\, - %
}\vspace{2mm}\\ %
\displaystyle{\hspace{06mm}%
  - \frac{2L}{\Lambda}\Psi_{1/2}\,d\mu\, -\frac{L}{\Lambda}\!\left[ 3\Psi_{3/2} + 2\tau\frac{\partial\Psi_{3/2}}{\partial\tau} \right]\!dT. %
}
\end{array}
\end{equation}
Here it was taken into account that %
$\partial\Psi_{3/2}\big/\partial\eta=\tau^{-1}\Psi_{1/2}$ and $d\varepsilon_R=-(2\varepsilon_R/R)dR$. %
Naturally, the usual thermodynamic relations hold  
$S=-\big(\partial\Omega\big/\partial T\big)_{\mu,L,R}$, \linebreak
$N=-\big(\partial\Omega\big/\partial \mu\big)_{T,L,R}$, as well as
$\Omega=E-TS-\mu N$.

\vspace{-0mm}
\section{Pressures} %
In a bulk system the pressure is connected with the thermodynamical
potential by the known formula \linebreak
$p=-\Omega/V$. In the considered
quasi-one-dimensional case the system is anisotropic since the
character of motion of particles along the tube and in the direction
of its radius is different and, therefore, the usual formula for
the pressure is invalid. The force exerted by the gas on the wall
perpendicular to the $z$ axis is different from the force exerted on
the side wall of the tube. These forces can be calculated in the
same way as in the volume case \cite{LL}. The pressures on the
planes perpendicular to the $z$ axis and on the wall of the tube, on
the basis of the relations $p =-(\partial E/\partial V)_S$ and $V=\pi R^2L$, %
are given by the formulas
\begin{equation} \label{14}
\begin{array}{ll}
\displaystyle{%
  p_\parallel =-\frac{1}{\pi R^2}\!\left( \frac{\partial E}{\partial L}\right)_{\!S,R}, \quad  %
  p_R =-\frac{1}{2\pi LR}\!\left( \frac{\partial E}{\partial R}\right)_{\!S,L},   %
}
\end{array}
\end{equation}
or
\begin{equation} \label{15}
\begin{array}{ll}
\displaystyle{%
  p_\parallel =-\frac{1}{\pi R^2}\!\left( \frac{\partial\Omega}{\partial L}\right)_{\!T,\mu,R}, \quad  %
  p_R =-\frac{1}{2\pi LR}\!\left( \frac{\partial\Omega}{\partial R}\right)_{\!T,\mu,L}.   %
}
\end{array}
\end{equation}
The differential of the thermodynamic potential (\ref{13}) can be
represented in the form
\begin{equation} \label{16}
\begin{array}{ll}
\displaystyle{%
  d\Omega= -p_\parallel\pi R^2dL - p_R2\pi LRdR - Nd\mu - SdT. %
}
\end{array}
\end{equation}
Considering the form of the thermodynamic potential (\ref{07}), we
obtain the formulas determining the pressures through the functions (\ref{06}): %
\begin{equation} \label{17}
\begin{array}{ll}
\displaystyle{\hspace{-0mm}%
  p_\parallel =\frac{2T}{\pi R^2\Lambda}\Psi_{3/2}, \quad  %
  p_R =\frac{2T}{\pi R^2\Lambda}\!\left( \frac{\eta}{\tau}\Psi_{1/2} + \tau\frac{\partial\Psi_{3/2}}{\partial\tau} \right) .   %
}
\end{array}
\end{equation}
The energy is connected with the pressures by the formula
\begin{equation} \label{18}
\begin{array}{ll}
\displaystyle{%
  E=\frac{1}{2}\pi R^2L\!\left( p_\parallel+2 p_R \right),  %
}
\end{array}
\end{equation}
which in the volume limit $p_\parallel=p_R=p$ \,turns into the known
relation $pV=(2/3)E$ \,for the Fermi gas \cite{LL}.

\section{Reduced form of thermodynamic quantities } %
It is convenient to introduce dimensionless quantities, which we
will call ``reduced'' and designate them by a tilde on top, for the
entropy, energy, pressures, volume and linear densities:
\begin{equation} \label{19}
\begin{array}{ll}
\displaystyle{%
  \tilde{S}\equiv\frac{2\pi^{1/2}}{3}\frac{R}{L}S, \qquad \tilde{E}\equiv 4\pi^{1/2}\frac{R^3}{L}\frac{m}{\hbar^2}E, %
}\vspace{2mm}\\ %
\displaystyle{\hspace{0mm}%
  \tilde{p}_\parallel\equiv 2\pi^{3/2}R^5\frac{m}{\hbar^2}p_\parallel, \qquad \tilde{p}_R\equiv 2\pi^{3/2}R^5\frac{m}{\hbar^2}p_R, %
}\vspace{2mm}\\ %
\displaystyle{\hspace{0mm}%
  \tilde{n}\equiv\pi^{3/2}R^3\,n, \qquad \tilde{n}_{\!L}\equiv\pi^{1/2}R\,n_{\!L}=\tilde{n}. %
}
\end{array}
\end{equation}
The reduced quantities are functions of only two independent
dimensionless variables -- the temperature $\tau$ and the chemical potential $\eta$:
\begin{equation} \label{20}
\begin{array}{ll}
\displaystyle{%
   \tilde{S}=\tau^{1/2}\!\left( \Psi_{3/2}+\frac{2}{3}\,\tau\frac{\partial\Psi_{3/2}}{\partial\tau} \right),  %
}\vspace{2mm}\\ %
\displaystyle{\hspace{0mm}%
  \tilde{E}=\tau^{3/2}\!\left( \Psi_{3/2}+ 2\frac{\eta}{\tau}\Psi_{1/2}+2\tau\frac{\partial\Psi_{3/2}}{\partial\tau} \right), %
}\vspace{2mm}\\ %
\displaystyle{\hspace{0mm}%
  \tilde{p}_\parallel=\tau^{3/2}\Psi_{3/2},\quad \tilde{p}_R=\tau^{3/2}\!\left(\frac{\eta}{\tau}\Psi_{1/2}+\tau\frac{\partial\Psi_{3/2}}{\partial\tau}\right), %
}\vspace{2mm}\\ %
\displaystyle{\hspace{0mm}%
  \tilde{n}=\tilde{n}_{\!L}=\tau^{1/2}\Psi_{1/2}. %
}
\end{array}
\end{equation}
The use of the reduces quantities is convenient owing to the fact
that they do not contain explicitly geometric dimensions of the system. %

\newpage
\section{Heat capacities }\vspace{-2mm} %
An important directly measurable thermodynamic quantity is the heat
capacity. In the geometry under consideration heat capacities can be
defined under various conditions different from those which take
place in the volume case. In order to obtain heat capacities, it is
necessary to calculate the quantity $C=T(dS/dT)$. For this purpose,
it is convenient to express the differential of the entropy through
the reduced quantities:
\begin{equation} \label{21}
\begin{array}{ll}
\displaystyle{\hspace{-3mm}%
   \frac{dS}{S}=  %
   \left(\frac{1}{\tilde{S}}\frac{\partial \tilde{S}}{\partial\eta}-\frac{1}{\tilde{n}_{\!L}}\frac{\partial \tilde{n}_{\!L}}{\partial\eta}\right)\!d\eta + \! %
   \left(\frac{1}{\tilde{S}}\frac{\partial \tilde{S}}{\partial\tau}-\frac{1}{\tilde{n}_{\!L}}\frac{\partial \tilde{n}_{\!L}}{\partial\tau}\right)\!d\tau. %
}
\end{array}
\end{equation}
In the volume case at the fixed number of particles, which is assumed
here, the equation of state is $p=p(T,V)$. If the chemical potential
is used as an independent variable, then the equation of state is
defined parametrically by the equations $p=p(T,V,\mu)$ and $N=N(T,V,\mu)$. %
To obtain the heat capacity as a function of only temperature, one
constraint should be imposed between the pressure and the volume. In
the simplest case, it its possible to fix either the volume or the
pressure, thus obtaining the heat capacities $C_V$ and $C_p$.

Under given conditions, owing to anisotropy of the system, there are
two equations of state (\ref{17}) for two pressures
$p_R=p_R(T,R,L,\mu)$ and $p_\parallel=p_\parallel(T,R,L,\mu)$, %
which at the fixed average number of particles should be considered
together with the equation $N=N(T,R,L,\mu)$. To obtain the heat
capacity as a function of only temperature, two additional
constraints should be set between the pressures $p_R, p_\parallel$
and the dimensions of the system $R,L$, namely
$F_1(p_R, p_\parallel,R,L)=0$ and $F_2(p_R, p_\parallel,R,L)=0$. %
In the simplest case, two of four quantities $p_R, p_\parallel,R,L$
can be fixed. Then the heat capacity as a function of temperature
can be considered under fixation of one of the following pairs of quantities: %
$(R,L)$, $(p_R, p_\parallel)$, $(R, p_\parallel)$, $(R, p_R)$, $(L, p_\parallel)$, $(L, p_R)$. %
Fixation of the first of pairs $(R,L)$ corresponds to the heat
capacity at a constant volume in the volume case, and of the second $(p_R, p_\parallel)$ -- \linebreak
to the heat capacity at a constant pressure. In calculations the
following relations should be taken into account: 
\begin{equation} \label{22}
\begin{array}{ll}
\displaystyle{%
   \frac{d\tilde{n}_{\!L}}{\tilde{n}_{\!L}}=\frac{dR}{R}-\frac{dL}{L}, \quad\,\,  %
   \frac{d\tau}{dT}=\frac{1}{\varepsilon_R}+\frac{2\tau}{R}\frac{dR}{dT}.
}
\end{array}
\end{equation}
Finally, we obtain the formulas for the reduced heat capacities
$\displaystyle{\tilde{C}\equiv\frac{2\pi^{1/2}}{3}\frac{R}{L}\,C}$
under different conditions, being valid at arbitrary temperatures: \vspace{-0mm} %
\vspace{0mm} %
\widetext
\begin{equation} \label{23}
\begin{array}{ll}
\displaystyle{%
   \tilde{C}_{LR}=\tau\!\left\{ \frac{\partial \tilde{S}}{\partial\tau}- %
   \frac{\partial \tilde{S}}{\partial\eta}\frac{\big(\partial\tilde{n}_{\!L}/\partial\tau\big)}{\big(\partial\tilde{n}_{\!L}/\partial\eta\big)} \right\},  %
}
\end{array}
\end{equation}
\begin{equation} \label{24}
\begin{array}{ll}
\displaystyle{%
   \tilde{C}_{p_\parallel p_R}=\tau\,
   \frac{
   \displaystyle{
   \left(\frac{\partial \tilde{S}}{\partial\eta}-\frac{\tilde{S}}{\tilde{n}_{\!L}}\frac{\partial \tilde{n}_{\!L}}{\partial\eta}\right)\!\!
   \left(\tilde{p}_R\frac{\partial \tilde{p}_\parallel}{\partial\tau}-\tilde{p}_\parallel\frac{\partial \tilde{p}_R}{\partial\tau}\right) + \! %
   \left(\frac{\partial \tilde{S}}{\partial\tau}-\frac{\tilde{S}}{\tilde{n}_{\!L}}\frac{\partial \tilde{n}_{\!L}}{\partial\tau}\right)\!\!
   \left(\tilde{p}_\parallel\frac{\partial \tilde{p}_R}{\partial\eta}-\tilde{p}_R\frac{\partial \tilde{p}_\parallel}{\partial\eta}\right) %
   }
   }
   {
   \displaystyle{
   \tilde{p}_\parallel\frac{\partial \tilde{p}_R}{\partial\eta}-\tilde{p}_R\frac{\partial \tilde{p}_\parallel}{\partial\eta} %
   +\frac{2}{5}\,\tau\!\left(\frac{\partial \tilde{p}_\parallel}{\partial\eta}\frac{\partial \tilde{p}_R}{\partial\tau}- %
   \frac{\partial \tilde{p}_\parallel}{\partial\tau}\frac{\partial \tilde{p}_R}{\partial\eta}\right) %
   }
   },
}
\end{array}
\end{equation}
\begin{equation} \label{25}
\begin{array}{ll}
\displaystyle{%
   \tilde{C}_{R p_\parallel}=\tau\!\left\{\!
   \left(\frac{\partial \tilde{S}}{\partial\tau}-\frac{\tilde{S}}{\tilde{n}_{\!L}}\frac{\partial \tilde{n}_{\!L}}{\partial\tau}\right)-\! %
   \left(\frac{\partial \tilde{S}}{\partial\eta}-\frac{\tilde{S}}{\tilde{n}_{\!L}}\frac{\partial \tilde{n}_{\!L}}{\partial\eta}\right)\!\! %
   \frac{\big(\partial\tilde{p}_\parallel/\partial\tau\big)}{\big(\partial\tilde{p}_\parallel/\partial\eta\big)} \right\},  %
}
\end{array}
\end{equation}
\begin{equation} \label{26}
\begin{array}{ll}
\displaystyle{%
   \tilde{C}_{L p_\parallel}=\tau\,
   \frac{
   \displaystyle{
   \frac{\tilde{n}_{\!L}}{2}\frac{\partial \tilde{p}_\parallel}{\partial\eta}\!
   \left(\frac{\partial \tilde{S}}{\partial\tau}-\frac{\tilde{S}}{\tilde{n}_{\!L}}\frac{\partial \tilde{n}_{\!L}}{\partial\tau}\right)\!- %
   \frac{\tilde{n}_{\!L}}{2}\frac{\partial \tilde{p}_\parallel}{\partial\tau}\!
   \left(\frac{\partial \tilde{S}}{\partial\eta}-\frac{\tilde{S}}{\tilde{n}_{\!L}}\frac{\partial \tilde{n}_{\!L}}{\partial\eta}\right)\!+ %
   \frac{5}{2}\,\tilde{p}_\parallel\!
   \left(\frac{\partial \tilde{S}}{\partial\eta}\frac{\partial \tilde{n}_{\!L}}{\partial\tau}-\frac{\partial \tilde{S}}{\partial\tau}\frac{\partial \tilde{n}_{\!L}}{\partial\eta}\right) %
   }
   }
   {
   \displaystyle{
   \frac{\tilde{n}_{\!L}}{2}\frac{\partial \tilde{p}_\parallel}{\partial\eta}-\frac{5}{2}\,\tilde{p}_\parallel\frac{\partial \tilde{n}_{\!L}}{\partial\eta}+ %
   \tau\!\left(\frac{\partial \tilde{n}_{\!L}}{\partial\eta}\frac{\partial \tilde{p}_\parallel}{\partial\tau}-\frac{\partial \tilde{n}_{\!L}}{\partial\tau}\frac{\partial \tilde{p}_\parallel}{\partial\eta}\right) %
   }
   }.
}
\end{array}
\end{equation}
\endwidetext\noindent
The heat capacities $\tilde{C}_{R p_R}$ and $\tilde{C}_{L p_R}$ are
determined by formulas (\ref{25}) and (\ref{26})
with account of the replacement $\tilde{p}_\parallel\rightarrow\tilde{p}_R$. %
Note that the formulas for the heat capacities
(\ref{23})\,--\,(\ref{25}) coincide with the corresponding heat
capacities of the Fermi gas enclosed between two planes %
under the replacement therein $R\rightarrow L_0$,
$\tilde{n}_{\!L}\rightarrow\tilde{n}_{\!A}$, where $L_0$ is the
distance between planes, $\tilde{n}_{\!A}$ is the reduced surface
density \cite{PS}. The heat capacity (\ref{26}) goes into the
formula for the two-dimensional case under the replacement
$\tilde{n}_{\!L}\rightarrow 2\tilde{n}_{\!A}$ \cite{PS}, manifesting
the quasi-one-dimensionality of the considered case.

\section{Compressibilities}
Another directly observable quantities are compressibilities. We
define the ``parallel'' and the ``radial'' compressibilities by the
relations
\begin{equation} \label{27}
\begin{array}{ll}
\displaystyle{%
   \gamma_\parallel = \frac{1}{n}\!\left( \frac{\partial n}{\partial p_\parallel}\right)_{\!R},\qquad  %
   \gamma_R = \frac{1}{n}\!\left( \frac{\partial n}{\partial p_R}\right)_{\!L}.  %
}
\end{array}
\end{equation}
The reduced compressibilities will be defined by the relation
$\tilde{\gamma}\equiv\big(\hbar^2\big/2\pi^{3/2}mR^5\big)\gamma$. %
Compressibilities can be calculated under the condition of both
constant temperature (isothermal) and constant entropy (adiabatic).
For the reduced compressibilities in isothermal conditions, we obtain: %
\begin{equation} \label{28}
\begin{array}{ll}
\displaystyle{%
   \tilde{\gamma}_{\parallel T} = %
   \frac{1}{\tilde{n}}\frac{\big(\partial \tilde{n}/\partial\eta\big)}{\big(\partial\tilde{p}_\parallel/\partial\eta\big)}, %
}
\end{array}
\end{equation}\vspace{-3mm}
\begin{equation} \label{29}
\begin{array}{ll}
\displaystyle{%
   \tilde{\gamma}_{R T} = %
  \frac{\big(\partial \tilde{n}/\partial\eta\big)}
  {\displaystyle{
  \left(\frac{5}{2}\,\tilde{p}_R-\tau\frac{\partial \tilde{p}_R}{\partial\tau}\right)\!\!\frac{\partial\tilde{n}}{\partial\eta} -\! %
  \left(\frac{\tilde{n}}{2}-\tau\frac{\partial \tilde{n}}{\partial\tau}\right)\!\!\frac{\partial \tilde{p}_R}{\partial\eta} } %
  }.
}
\end{array}
\end{equation}

The adiabaticity condition consists in the invariance of the entropy
per one particle (and therefore of the total entropy in the system
with a fixed number of particles). In the considered case the
adiabaticity condition has the form:
\begin{equation} \label{30}
\begin{array}{ll}
\displaystyle{%
   \sigma\equiv \frac{S}{N}= \frac{1}{\Psi_{1/2}}\!\left(\frac{3}{2}\Psi_{3/2}+\tau\frac{\partial\Psi_{3/2}}{\partial\tau} \right)\!\equiv  %
   \Theta(\tau,\eta)=\rm{const}.
}
\end{array}
\end{equation}
Together with the equation for the number of particles (\ref{08}),
the equation (\ref{30}) determines relationships between the
density, temperature and pressures in adiabatic processes. The
adiabatic compressibilities are given by the formulas:
\begin{equation} \label{31}
\begin{array}{ll}
\displaystyle{%
\tilde{\gamma}_{\parallel \sigma} = %
  \frac{
  \displaystyle{
  \frac{\partial\Theta}{\partial\eta}\frac{\partial\tilde{n}}{\partial\tau} - \frac{\partial\Theta}{\partial\tau}\frac{\partial\tilde{n}}{\partial\eta} %
  }
  }
  {\displaystyle{
  \tilde{n}\!\left(\!\frac{\partial\Theta}{\partial\eta}\frac{\partial\tilde{p}_\parallel}{\partial\tau} - \frac{\partial\Theta}{\partial\tau}\frac{\partial\tilde{p}_\parallel}{\partial\eta}\!\right)} %
  },
}
\end{array}
\end{equation}\vspace{-3mm}
\begin{equation} \label{32}
\begin{array}{ll}
\displaystyle{%
\tilde{\gamma}_{R \sigma} %
  \!=\!\frac{
  \displaystyle{
  2\!\left(\!\frac{\partial\Theta}{\partial\eta}\frac{\partial\tilde{n}}{\partial\tau} - \frac{\partial\Theta}{\partial\tau}\frac{\partial\tilde{n}}{\partial\eta}\!\right) %
  }
  }
  {\displaystyle{
  5\tilde{p}_R\!\left(\!\frac{\partial\Theta}{\partial\eta}\frac{\partial \tilde{n}}{\partial\tau} -\! %
  \frac{\partial\Theta}{\partial\tau}\frac{\partial\tilde{n}}{\partial\eta}\!\right)\! -
  \tilde{n}\!\left(\!\frac{\partial\Theta}{\partial\eta}\frac{\partial\tilde{p}_R}{\partial\tau} -\! %
  \frac{\partial\Theta}{\partial\tau}\frac{\partial\tilde{p}_R}{\partial\eta}\!\right)
  } %
  }.
}
\end{array}
\end{equation}
Certainly, at zero temperature the isothermal and adiabatic
compressibilities coincide.

\section{Analysis of functions \newline $\Psi_{1/2}(\tau,\eta)$ and $\Psi_{3/2}(\tau,\eta)$ } %
As shown above, all thermodynamic quantities are expressed through
the functions $\Psi_{1/2}(\tau,\eta)$, $\Psi_{3/2}(\tau,\eta)$ and their
derivatives. In this section we study some properties of these
functions. Note that when studying oscillations in the Fermi gas with quantized
levels, usually the Poisson formula is used for the extraction of the
oscillating part of thermodynamic and kinetic quantities \cite{LL}. But a detailed analysis
undertaken by the authors shows that it is more convenient to
calculate the functions, by which thermodynamic quantities are expressed,
without use of the Poisson formula. This, in particular, is connected
with the fact that the possibility of extraction of an oscillating part
in some function does not at all mean that the total function is oscillating,
and the contribution of non-oscillating part should be analyzed as well \cite{PS}.

At fixed particle number density and at high temperatures, the same
as in the volume case, the chemical potential is negative. With
decreasing temperature it increases and at some temperature $T_0$
turns into zero $(\eta=0)$, becoming positive at $T<T_0$. There is one
more characteristic temperature $T_R$, at which %
$\mu=\rho_{01}^2\varepsilon_R$ ($\eta=\rho_{01}^2$), %
$\rho_{01}$ is the lowest zero of the Bessel function $J_n(\rho_{n\nu})=0$. %
The dependencies of the dimensionless chemical potential
$\eta$ on the dimensionless temperature $\tau$ are shown in Fig.\,1.
The characteristic temperatures $\tau_0=T_0/\varepsilon_R$ and $\tau_R=T_R/\varepsilon_R$
are determined from the equations:
\begin{equation} \label{33}
\begin{array}{ll}
\displaystyle{%
   \tilde{n}_{\!L}=\tau_0^{1/2}\Psi_{1/2}(\tau_0,0),\quad \tilde{n}_{\!L}=\tau_R^{1/2}\Psi_{1/2}(\tau_R,1). %
}
\end{array}
\end{equation}
The region where $\mu<\rho_{01}^2\varepsilon_R$ ($\eta<\rho_{01}^2$)
will be for convenience called the high temperature region, and the region
$\mu>\rho_{01}^2\varepsilon_R$ ($\eta>\rho_{01}^2$) -- the low temperature region.
\vspace{0mm} %
\begin{figure}[h!]
\vspace{0mm} \hspace{-0mm}
\includegraphics[width = 0.99\columnwidth]{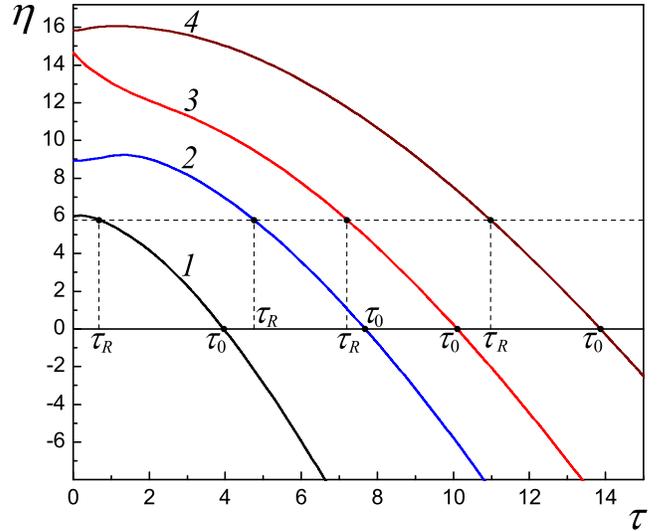} 
\vspace{-8mm} 
\caption{\label{fig01} %
The dependencies of the chemical potential on temperature $\eta(\tau)$
at different values of the reduced density: \newline %
({\it 1}) $\tilde{n}_{\!L}=0.5$ ($\tau_R=0.68$, $\tau_0=3.96$); %
({\it 2}) $\tilde{n}_{\!L}=2.0$ ($\tau_R=4.76$, $\tau_0=7.68$); %
({\it 3}) $\tilde{n}_{\!L}=(2/\sqrt{\pi})\big(\rho_2^2-\rho_1^2\big)^{1\!/2}=3.37$ ($\tau_R=7.20$, $\tau_0=10.10$); %
({\it 4}) $\tilde{n}_{\!L}=6.0$ ($\tau_R=10.98$, $\tau_0=13.87$). %
%
}%
\end{figure}

It is convenient to arrange the zeros of the Bessel function in ascending order
with the help of a single new index $r\equiv(n,\nu)$, so that $\rho_{r+1}>\rho_r$, 
where $1\equiv(0,1)$ corresponds to the lowest value at which the Bessel function
becomes zero $\rho_1=\rho_{01}$. The inequalities hold \cite{AS}:
\begin{equation} \label{34}
\begin{array}{ll}
\displaystyle{\hspace{-3mm}%
   \rho_{n1}<\rho_{n+1,1}<\rho_{n2}<\rho_{n+1,2}<\rho_{n3}<\rho_{n+1,3}<\ldots\,.\hspace{-2mm}  %
}
\end{array}
\end{equation}
In Table I the ten first zeros of the Bessel function arranged
in ascending order are given for reference. 
In this notation, the functions (\ref{06}) are written in the form
\begin{equation} \label{35}
\begin{array}{ll}
\displaystyle{%
  \Psi_s(\tau,\eta)\equiv\sum_{r=1}^\infty g_r\Phi_s\!\left[\tau^{-1}\big(\eta-\rho_r^2\big)\right], %
}
\end{array}
\end{equation}
where $g_r=1$ at $n=0$ and $g_r=2$ at $n\geq 1$.
\vspace{-4mm}%
\begin{table}[h!] \nonumber
\centering %
\caption{The ten first zeros of the Bessel function } %
\vspace{0.5mm}%
\begin{tabular}{|c|c|c|c|c|} \hline  
\,$\rho_1=\rho_{0,1}$  & \,$\rho_2=\rho_{1,1}$ & \,$\rho_3=\rho_{2,1}$ & \,$\rho_4=\rho_{0,2}$ & \,$\rho_5=\rho_{3,1}$    \\ \hline %
2.40                   &  3.83                 &  5.14                 &  5.52                 &  6.38                    \\ \hline %
\,$\rho_6=\rho_{1,2}$  & \,$\rho_7=\rho_{4,1}$ & \,$\rho_8=\rho_{2,2}$ & \,$\rho_9=\rho_{0,3}$ & \,$\rho_{10}=\rho_{5,1}$ \\ \hline %
7.02                   &  7.59                 &  8.42                 &  8.65                 &  8.77                    \\ \hline %
\end{tabular}  
\end{table}

In the high temperature region $\eta\leq\rho_1^2$ the functions
(\ref{06}) can be calculated by the formula
\begin{equation} \label{36}
\begin{array}{ll}
\displaystyle{%
   \Psi_s(\tau,\eta)=\sum_{k=1}^\infty\!
   \frac{(-1)^{k+1}}{k^s}\sum_{r=1}^\infty g_r\,e^{k (\eta-\rho_r^2)\!/\tau} %
}
\end{array}
\end{equation}
In the low temperature region $\eta > \rho_1^2$ there can be obtained
asymptotic expansions for the functions (\ref{06}), valid in the limit $\tau\rightarrow 0$.
Let us represent the functions (\ref{06}) in the form
\begin{equation} \label{37}
\begin{array}{ll}
\displaystyle{%
  \Psi_s(\tau,\eta)=
}\vspace{2mm}\\ %
\displaystyle{\hspace{1mm}%
  =\sum_{r=1}^{r_0-1}\! g_r\Phi_s\!\left[\tau^{-1}\big(\eta-\rho_r^2\big)\right] %
                   +\sum_{r=r_0+2}^\infty\! g_r\Phi_s\!\left[\tau^{-1}\big(\eta-\rho_r^2\big)\right] + %
}\vspace{2mm}\\ %
\displaystyle{\hspace{1mm}%
  +\,g_{r_0}\Phi_s\!\left[\tau^{-1}\big(\eta-\rho_{r_0}^2\big)\right] %
  +g_{r_0+1}\Phi_s\!\left[\tau^{-1}\big(\eta-\rho_{r_0+1}^2\big)\right]. %
}
\end{array}
\end{equation}
Here the index $r_0$, enumerating zeros, is defined by the condition:
$\rho_{r_0}^2<\eta<\rho_{r_0+1}^2$.
We singled out the third and the fourth terms,
important at $\eta\geq\rho_{r_0}^2$ and $\eta\leq\rho_{r_0+1}^2$. %
At $\tau\ll 1$ the contribution of the second term is exponentially small,
and for calculation of the first sum one should make use of the expansion
being valid to within exponential corrections:
\begin{equation} \label{38}
\begin{array}{l}
\displaystyle{%
  \Phi_s(t)=\frac{t^s}{s\Gamma(s)}\times %
}\vspace{2mm}\\ %
\displaystyle{\hspace{0mm}%
  \times\!\left\{1+2\sum_{k=0}^{\infty}\!\frac{s(s\!-\!\!1)\ldots(s\!-\!\!2k\!-\!\!1)(1\!-\!2^{\,-2k-\!1})\zeta(2k\!+\!2)}{t^{2k+2}} \!\right\}\!. %
}
\end{array}
\end{equation}
Finally, we obtain the asymptotic expansion valid in the limit $\tau\rightarrow 0$:
\begin{equation} \label{39}
\begin{array}{ll}
\displaystyle{%
  \Psi_s(\tau,\eta)=\frac{1}{s\Gamma(s)\,\tau^s}\,Z_s^{(r_0-1)}(\eta)\,+
}\vspace{2mm}\\ %
\displaystyle{\hspace{-0mm}%
  +\frac{2}{\Gamma(s)}\!\sum_{k=0}^{\infty}\! b_s(2k\!+\!1\!)(1\!\!-\!2^{-2\!k-\!1}\!)\zeta(2k\!+\!2)\tau^{2\!k+2\!-\!s}Z_{s-2\!k-\!2}^{(r_0-1)}(\eta)+ %
}\vspace{2mm}\\ %
\displaystyle{\hspace{0mm}%
  +\,g_{r_0}\Phi_s\!\left[\tau^{-1}\big(\eta-\rho_{r_0}^2\big)\right] %
  +g_{r_0+1}\Phi_s\!\left[\tau^{-1}\big(\eta-\rho_{r_0+1}^2\big)\right], %
}
\end{array}
\end{equation}
where
\begin{equation} \label{40}
\begin{array}{ll}
\displaystyle{%
  Z_s^{(r_0)}(\eta)\equiv\sum_{r=1}^{r_0} g_r\big(\eta-\rho_r^2\big)^s, %
}
\end{array}
\end{equation}
$b_s(m)=\prod_{l=1}^m(s-l)$ at $m\geq 1$, and $b_s(0)=1$.
In particular, for the functions required in calculation of thermodynamic quantities,
at $\tau\rightarrow 0$ it is sufficient to use the following approximations:
\begin{equation} \label{41}
\begin{array}{ll}
\displaystyle{%
  \Psi_{1/2}(\tau,\eta)=\frac{2}{(\pi\tau)^{1/2}}\,Z_{1/2}^{(r_0-1)}(\eta)\,+
}\vspace{2mm}\\ %
\displaystyle{\hspace{1mm}%
  +\,g_{r_0}\Phi_{1/2}\!\left[\tau^{-1}\big(\eta-\rho_{r_0}^2\big)\right] %
  +g_{r_0+1}\Phi_{1/2}\!\left[\tau^{-1}\big(\eta-\rho_{r_0+1}^2\big)\right], %
}\vspace{3.8mm}\\ %
\displaystyle{%
  \Psi_{3/2}(\tau,\eta)=\frac{4}{3\sqrt{\pi}\tau^{3/2}}\left[Z_{3/2}^{(r_0-1)}(\eta)+\frac{\pi^2}{8}\,\tau^2Z_{-1/2}^{(r_0-1)}(\eta)\right]\!+
}\vspace{2mm}\\ %
\displaystyle{\hspace{1mm}%
  +\,g_{r_0}\Phi_{3/2}\!\left[\tau^{-1}\big(\eta-\rho_{r_0}^2\big)\right] %
  +g_{r_0+1}\Phi_{3/2}\!\left[\tau^{-1}\big(\eta-\rho_{r_0+1}^2\big)\right]. %
}\
\end{array}
\end{equation}
In fact, at $\tau\leq 1$ these formulas approximate well
the exact functions defined by the formula (\ref{06}).
Thus, at $\tau=0.1$ the error does not exceed 0.01\%, and even at $\tau=1$ -- 1\%. %
The dependencies of these functions and their derivatives
on the chemical potential are shown in Fig.\,2.
As seen (Fig.\,2b), the function $\Psi_{3/2}(\tau,\eta)$ and its derivative
monotonically increase with increasing the chemical potential.
The function $\Psi_{1/2}(\tau,\eta)$ also monotonically increases
but the behavior of its derivative has an oscillating character,
besides the derivative has sharp maximums at the points corresponding to
the zeros of the Bessel function (Fig.\,2a). Note also that,
since $\tilde{n}=\tilde{n}_{\!L}=\tau^{1/2}\Psi_{1/2}(\tau,\eta)$, %
the function $\Psi_{1/2}(\tau,\eta)$ describes the dependence of
the reduced density, divided by $\tau^{1/2}$, on the chemical \linebreak potential.
%
\begin{figure*}[t!]
\vspace{0mm} \hspace{-0mm}
\includegraphics[width = 1.0\textwidth]{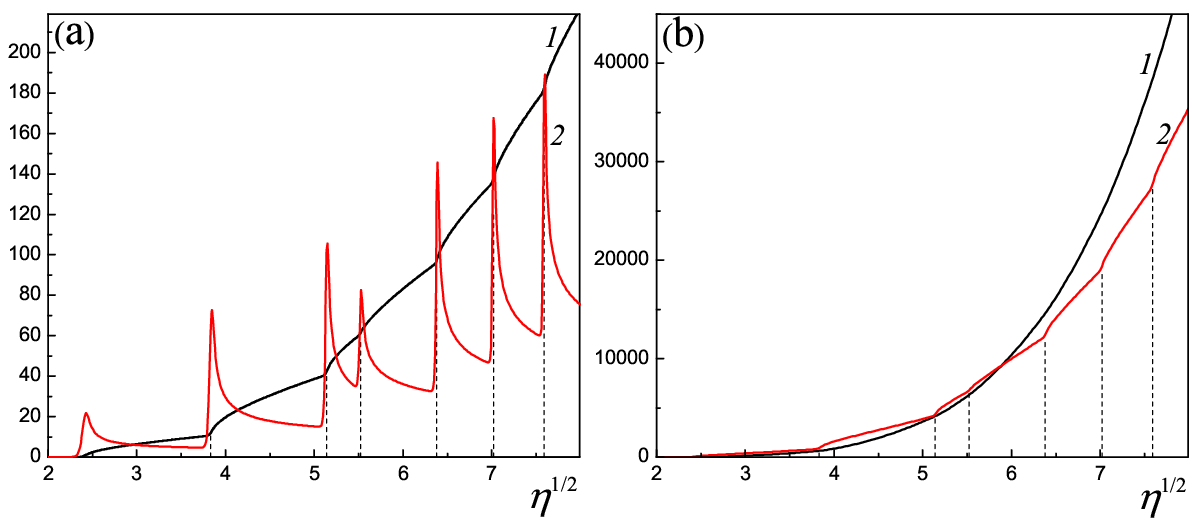} 
\vspace{-6mm}
\caption{\label{fig02} %
Graphs of the functions $\Psi_{1/2}(\tau,\eta)$, $\Psi_{3/2}(\tau,\eta)$ and their derivatives at $\tau=0.1$. \newline %
(a) The functions $\Psi_{1/2}(\tau,\eta)$ ({\it 1}) and $\partial\Psi_{1/2}(\tau,\eta)\big/\partial\eta^{1/2}$ ({\it 2}); %
(b) The functions $\Psi_{3/2}(\tau,\eta)$ ({\it 1}) and $\partial\Psi_{3/2}(\tau,\eta)\big/\partial\eta^{1/2}$ ({\it 2}). \newline %
The dashed lines correspond to the zeros of the Bessel function $\eta^{1/2}=\rho_r$.
}%
\end{figure*}

\section{Thermodynamic quantities \newline at low temperatures } %
The most interesting range where the quantum effects may manifest themselves
on the macroscopic level in the behavior of thermodynamic characteristics
is the range of low temperatures. Let us consider the behavior
of observable quantities at low temperatures such that \linebreak $\tau\ll 1$
using the formulas (\ref{41}). For the reduces densities of particles we have 
\begin{equation} \label{42}
\begin{array}{ll}
\displaystyle{%
  \tilde{n}=\tilde{n}_{\!L}=\tau^{1/2}\Psi_{1/2}(\tau,\eta)=\frac{2}{\sqrt{\pi}}\,Z_{1/2}^{(r_0-1)}(\eta)\,+
}\vspace{2mm}\\ %
\displaystyle{\hspace{-1mm}%
  +\tau^{1/2}\Big\{g_{r_0}\Phi_{1\!/2}\!\!\left[\tau^{-1}\!\big(\eta\!-\!\rho_{r_0}^2\!\big)\right] %
  \!+g_{r_0+1}\Phi_{1\!/2}\!\!\left[\tau^{-1}\!\big(\eta\!-\!\rho_{r_0+1}^2\!\big)\right]\!\!\Big\}\!. %
}
\end{array}
\end{equation}
Here with a good accuracy the density can be considered independent of temperature and the formula  %
$\tilde{n}=\tilde{n}_{\!L}\approx (2/\!\sqrt{\pi})Z_{1/2}^{(r_0)}(\eta)$ can be used. %
The reduced entropy is determined by the formula \newline\vspace{-2mm}
\begin{equation} \label{43}
\begin{array}{ll}
\displaystyle{%
  \tilde{S}=\frac{2\pi^{3/2}}{9}\,Z_{-1/2}^{(r_0-1)}(\eta)\,\tau\,+
}\vspace{2mm}\\ %
\displaystyle{\hspace{-1mm}%
  +\,\tau^{1/2}\Big\{g_{r_0}\Phi_{3\!/2}\!\!\left[\tau^{-1}\!\big(\eta\!-\!\rho_{r_0}^2\!\big)\right] %
  \!+g_{r_0+1}\Phi_{3\!/2}\!\!\left[\tau^{-1}\!\big(\eta\!-\!\rho_{r_0+1}^2\!\big)\right]\!- 
}\vspace{2mm}\\ %
\displaystyle{\hspace{12mm}%
  -(2/3)\tau^{-1}\!\big(\eta\!-\!\rho_{r_0}^2\!\big)g_{r_0}\Phi_{1\!/2}\!\!\left[\tau^{-1}\!\big(\eta\!-\!\rho_{r_0}^2\!\big)\right]- %
}\vspace{2mm}\\ %
\displaystyle{\hspace{12mm}%
  -(2/3)\tau^{-1}\!\big(\eta\!-\!\rho_{r_0+1}^2\!\big)g_{r_0+1}\Phi_{1\!/2}\!\!\left[\tau^{-1}\!\big(\eta\!-\!\rho_{r_0+1}^2\!\big)\right]\!\Big\}. %
}
\end{array}
\end{equation}
The terms which contain the functions $\Phi_s$ in (\ref{43})
can be significant when $\eta\approx\rho_{r_0}^2$ or $\eta\approx\rho_{r_0+1}^2$. %
It should be noted that in the quasi-one-dimensional case,
the same as in the volume and the quasi-two-dimensional \cite{PS} cases,
nearly everywhere except the specific points there remains
the linear dependence of the entropy on temperature (see Fig.\,3),
and the following formula can be used
\begin{equation} \label{44}
\begin{array}{ll}
\displaystyle{%
  \tilde{S}\approx\frac{2\pi^{3/2}}{9}\,Z_{-1/2}^{(r_0)}(\eta)\,\tau.
}
\end{array}
\end{equation}
The slope of the lines (\ref{44}) and the width of the range
where this formula is valid depend on the closeness of
the chemical potential value to the square of the Bessel function zero.
As the chemical potential approaches the square of the Bessel function zero
from the side of greater values $\eta\rightarrow\rho_{r_0}^2+0$
the slope angle of a line increases (curve 2 in Fig.\,3) and
at the very point $\eta=\rho_{r_0}^2$, as it follows from the formula (\ref{43}),
$\tilde{S}\sim\sqrt{\tau}$ (curve 1 in Fig.\,3).
As the chemical potential approaches the square of the Bessel function zero
from the side of smaller values $\eta\rightarrow\rho_{r_0}^2-0$ %
the slope angle of a line tends to a finite value (curve 3 in Fig.\,3),
at that, however, the temperature range where the linear dependence is realized
proves to be quite narrow.

The dependence of the reduced entropy on the dimensionless
chemical potential is presented in Fig.\,4.
As we see, the entropy has sharp maximums at the points
where the chemical potential equals the square of the Bessel function zero.
At calculation by means of the approximate formula (\ref{44})
while approaching the point $\eta=\rho_{r_0}^2$ from the right
there is the root singularity
$\tilde{S}\sim\big(\eta-\rho_{r_0}^2\big)^{\!-1\!/2}$,
and while approaching this point from the left the entropy takes a finite value.
The calculation by the more accurate formula (\ref{43}) eliminates the singularity
and allows to obtain the entropy value at the maximum $\tilde{S}_{\rm{max}}$, %
which with a good accuracy is determined by the formula
\begin{equation} \label{45}
\hspace{00mm}
\begin{array}{ll}
\displaystyle{%
  \tilde{S}_{\rm{max}}\approx\tilde{S}(\tau,\eta_{r_0})\approx
}\vspace{2mm}\\ %
\displaystyle{\hspace{03mm}%
  \approx\!\frac{2\pi^{3/2}}{9}\,Z_{-1/2}^{(r_0-1)}(\eta_{r_0})\,\tau +
  g_{r_0}\!\big(1\!-\!2^{-1/2}\big)\zeta(3/2)\sqrt{\tau}, %
}
\end{array}
\end{equation}
where $\zeta(3/2)\approx 2.612$ is the Riemann zeta function.
Pay attention that the values of maximums for the nondegenerate levels with $n=0$\,
turn out to be less than the appropriate values for the neighboring levels with
the two-fold degeneracy at $n\neq 0$ (see Fig.\,4).

In the quasi-one-dimensional case under study the dependence of
the entropy on the chemical potential is qualitatively different from
the quasi-two-dimensional case \cite{PS}, i.e. the entropy has a sharp maximum
rather than undergoes a jump at the beginning of the filling of a new discrete level.
One more difference of the given system from the quasi-two-dimensional case is
the irregularity of the values of chemical potential, specified
by the Bessel function zeros, at which the entropy has peculiarities \cite{PS}.

\vspace{0mm} %
\begin{figure}[t!]
\vspace{0mm} \hspace{-0mm}
\includegraphics[width = 1.0\columnwidth]{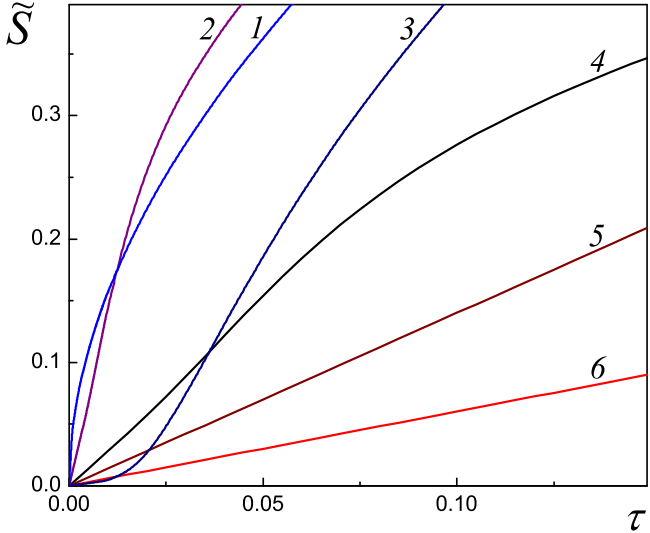} 
\vspace{-8mm} 
\caption{\label{fig03} %
The dependence of the reduced entropy $\tilde{S}(\tau,\eta)$ on the temperature $\tau$ %
at a fixed value of the chemical potential: \newline %
({\it 1}) $\eta=\rho_2^2=\rho_{1,1}^2=14.68$ ($\tilde{n}_{\!L}=3.37$); %
({\it 2}) $\eta=14.72$ ($\tilde{n}_{\!L}=3.81$); %
({\it 3}) $\eta=14.60$ ($\tilde{n}_{\!L}=3.35$); %
({\it 4}) $\eta=5.98$ ($\tilde{n}_{\!L}=0.5$); \newline 
({\it 5}) $\eta=20.0$ ($\tilde{n}_{\!L}=9.46$); %
({\it 6}) $\eta=10.0$ ($\tilde{n}_{\!L}=2.32$). \newline %
The density $\tilde{n}_{\!L}$ is given at $\tau=0$.
}%
\end{figure}

\newpage\vspace{-0mm}
In the main approximation all introduced above heat capacities
(\ref{23})\,--\,(\ref{26}), as it had to be expected, prove to be
identical and coinciding with the entropy.
Thus, in the quasi-one-dimensional case,
the same as in the quasi-two-dimensional \cite{PS} and the volume cases,
all heat capacities prove to be proportional to temperature:
\begin{equation} \label{46}
\begin{array}{ll}
\displaystyle{%
  \tilde{C}=\frac{2\pi^{3/2}}{9}\,Z_{-1/2}^{(r_0)}(\eta)\,\tau.
}
\end{array}
\end{equation}
Only at the specific points, at $\eta\rightarrow\rho_{r_0}^2+0$,
the same as in the entropy case, there appears the dependence $\tilde{C}\sim\sqrt{\tau}$.
The temperature dependence of heat capacities and their dependence
on the chemical potential are similar to the dependencies for the entropy shown in Figs.\,\,3 and 4.

It is interesting to consider the transition from the formulas
(\ref{44}), (\ref{46}) to the volume case, when the radius of the tube
becomes large as compared with the de Broglie wavelength.
It can be done, if one takes into account that in order to make the transition from
the dispersion law of particles in the tube (\ref{02}) to the dispersion law of free particles
it is necessary to make the substitution
\begin{equation} \label{47}
\begin{array}{l}
\displaystyle{%
  \rho_{n\nu}^2\frac{\hbar^2}{2mR^2} \,\rightarrow\, \frac{\hbar^2(k_x^2+k_y^2)}{2m}. %
}%
\end{array}
\end{equation}
The same rule of transition to the continuous spectrum can be obtained
by means of the asymptotic expression for the Bessel functions.

Performing formally the same substitution at calculation of the functions (\ref{41})
and replacing the summation over the Bessel function zeros
by the integration over the wave vector in the $(x,y)$ plane, we obtain
\begin{equation} \label{48}
\begin{array}{ll}
\displaystyle{%
  Z_{1/2}(\eta)=\frac{\sqrt{2}(m\mu)^{3/2}R^3}{3\hbar^3},\quad %
  Z_{-1/2}(\eta)=\frac{(m\mu)^{1/2}R}{\sqrt{2}\hbar}. %
}%
\end{array}
\end{equation}
Finally, we come to the known volume expressions for
the particle number density $n=(2m\mu)^{3/2}\!\big/3\pi^2\hbar^3$
and also the entropy and the heat capacity
$S=C=\big(m^{3/2}(2\mu)^{1/2}\!\big/3\hbar^3\big)VT$. %
%
\vspace{0mm} %
\begin{figure}[t!]
\vspace{0mm} \hspace{-0mm}
\includegraphics[width = 1.01\columnwidth]{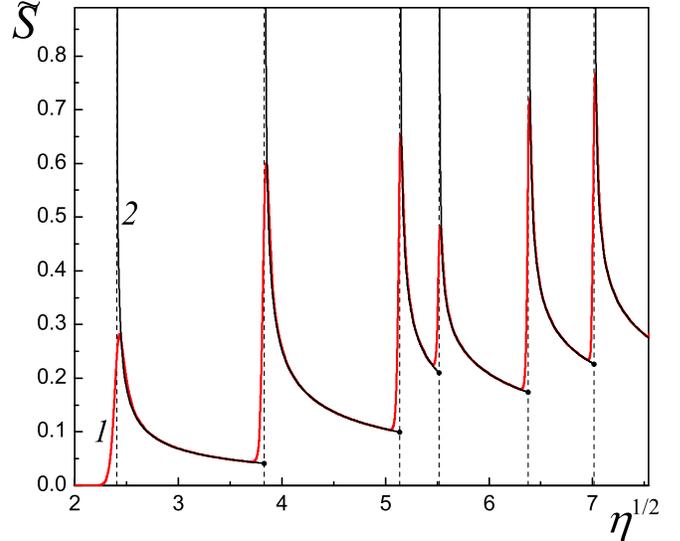} 
\vspace{-8mm} 
\caption{\label{fig04} %
The dependence of the reduced entropy $\tilde{S}(\tau,\eta)$ on the chemical potential. 
The curve ${\it 1}$ is plotted by means of the formula (43), and the curve ${\it 2}$ -- by the formula (44) at $\tau=0.1$. %
The dashed lines correspond to the zeros of the Bessel function $\eta^{1/2}=\rho_r$.
}%
\end{figure}

Let us give the results of calculation of the reduced pressures at zero temperature.
In this case the pressures are determined by the formulas:
\begin{equation} \label{49}
\begin{array}{ll}
\displaystyle{%
  \tilde{p}_\parallel=\frac{4}{3\sqrt{\pi}}\,Z_{3/2}^{(r_0)}(\eta),%
}%
\end{array}
\end{equation}\vspace{-5mm}
\begin{equation} \label{50}
\begin{array}{ll}
\displaystyle{%
  \tilde{p}_R=\frac{2}{\sqrt{\pi}}\Big[\eta Z_{1/2}^{(r_0)}(\eta)-Z_{3/2}^{(r_0)}(\eta)\Big]. %
}%
\end{array}
\end{equation}
The dependencies of the pressures (\ref{49}) and  (\ref{50}) on the chemical potential
are shown in Fig.\,5. As seen, the parallel pressure smoothly increases
with increasing the chemical potential, and the radial pressure undergoes breaks
(discontinuities in the derivative) at the points
where the filling of a new discrete level begins.
In the volume case, considering the formula
$Z_{3/2}(\eta)=2^{3/2}(m\mu)^{5/2}R^5\!\big/5\hbar^5$, 
the both formulas (\ref{49}) and (\ref{50}) lead to the known expression
for the pressure of the Fermi gas at zero temperature:
$ p=\big(2^{5/2}\!\big/15\pi^2\big)\,m^{3/2}\mu^{5/2}\!\big/\hbar^3. $ %
\vspace{0mm} %
\begin{figure}[t!]
\vspace{0mm} \hspace{-0mm}
\includegraphics[width = 1.0\columnwidth]{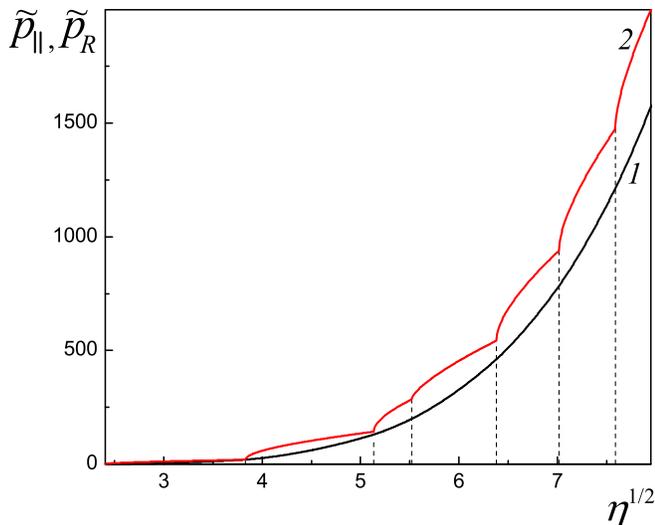} %
\vspace{-8mm} 
\caption{\label{fig05} %
The dependencies of the reduced pressures on the chemical potential at zero temperature:
({\it 1}) $\tilde{p}_\parallel(\eta)$; ({\it 2}) $\tilde{p}_R(\eta)$. %
Dashed lines correspond to the zeros of the Bessel function. %
}%
\end{figure}

Let us proceed to consideration of the compressibilities.
The reduced compressibilities at zero temperature are determined be the formulas:
\begin{equation} \label{51}
\begin{array}{ll}\hspace{0mm}
\displaystyle{%
  \tilde{\gamma}_\parallel=\frac{\sqrt{\pi}}{4}\frac{Z_{-1/2}^{(r_0)}(\eta)}{\big[Z_{1/2}^{(r_0)}(\eta)\big]^2}, %
}\vspace{1mm}\\ %
\displaystyle{\hspace{00mm}%
   \tilde{\gamma}_R=\!\frac{\sqrt{\pi}\,Z_{-1\!/2}^{(r_0)}(\eta)}
   {4\eta Z_{1\!/2}^{(r_0)}\!(\eta)Z_{-1\!/2}^{(r_0)}\!(\eta)\!-\!5 Z_{3\!/2}^{(r_0)}\!(\eta)Z_{-1\!/2}^{(r_0)}\!(\eta)\!+\!\!\big[Z_{1\!/2}^{(r_0)}\!(\eta)\big]^2}. 
}
\end{array}
\end{equation}
The dependencies of the compressibilities (\ref{51}) on the chemical potential are shown in Fig.\,6.
While approaching the point where the filling of a new level begins
from the side of smaller values of the chemical potential
the parallel compressibility has a finite value,
and while approaching the specific point
from the side of greater values of the chemical potential -- it tends to infinity.
The radial compressibility has finite maximums in the specific points, such that
the magnitude of peaks rapidly decreases with increasing the chemical potential.
In the volume limit the both compressibilities coincide, being given by the formula
\begin{equation} \label{52}
\begin{array}{ll}\hspace{0mm}
\displaystyle{%
  \gamma_\parallel=\gamma_R=\frac{9\pi^2}{4\sqrt{2}}\frac{\hbar^3}{m^{3/2}\mu^{5/2}}. %
}
\end{array}
\end{equation}
Since the square of speed of sound $u^2=m^{-1}\big(\partial p/\partial n\big)$ %
and the compressibility are connected by the relation $u^2=1\big/ mn\gamma$, %
then from (\ref{52}) it follows the known formula for the speed of sound
in the Fermi gas at zero temperature: $u^2=(2/3)\,\mu/m$. %
\vspace{0mm} %
\begin{figure}[t!]
\vspace{0mm} \hspace{-0mm}
\includegraphics[width = 1.0\columnwidth]{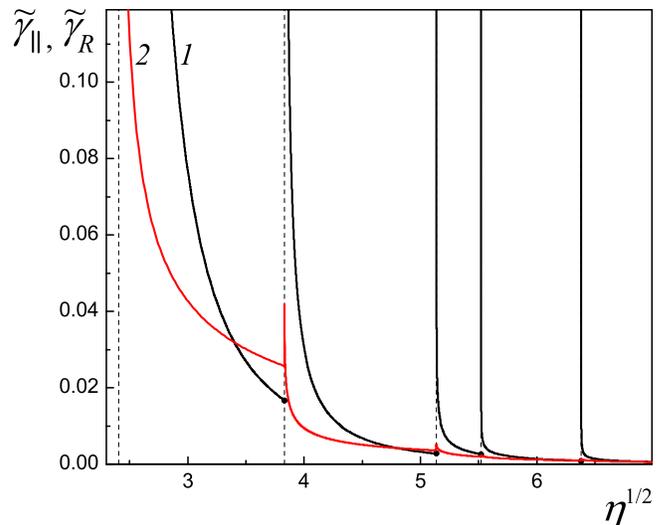} %
\vspace{-8mm} 
\caption{\label{fig06} %
The dependencies of the compressibilities on the chemical potential at zero temperature: 
({\it 1}) $\tilde{\gamma}_\parallel(\eta)$; ({\it 2}) $\tilde{\gamma}_R(\eta)$. %
Dashed lines correspond to the zeros of the Bessel function. %
}%
\end{figure}

The peculiarities considered above in the behavior of
the entropy, heat capacities, pressures and compressibilities of the quasi-one-dimensional system
and their difference from those of the quasi-two-dimensional system are connected with the form of the density of states $\nu(\varepsilon)$,
which is shown in Fig.\,7 for the quasi-two-dimensional quantum well and the quasi-one-dimensional tube.
In the case of the quantum well the density of states is constant and changes by a jump at the beginning of the filling of a new level (Fig.\,7a).
In the quasi-one-dimensional case the density of states tends to infinity
$\nu(\varepsilon)\sim\big(\varepsilon-\rho_r^2\varepsilon_R\big)^{\!-1/2}$ at $\varepsilon\rightarrow\rho_r^2\varepsilon_R+0$ (Fig.\,7b).
Such a dependence of the density of states on the energy is similar to that which takes place for electrons in magnetic field.
The difference consists in that in magnetic field the peculiarities are distributed regularly with the intervals multiple of odd numbers
of half the cyclotron frequency. In the considered case the distribution of the peculiarities is irregular and is determined by the zeros of the Bessel function.
%
\begin{figure*}[t!]
\vspace{-5mm} \hspace{-0mm}
\includegraphics[width = 1.0\textwidth]{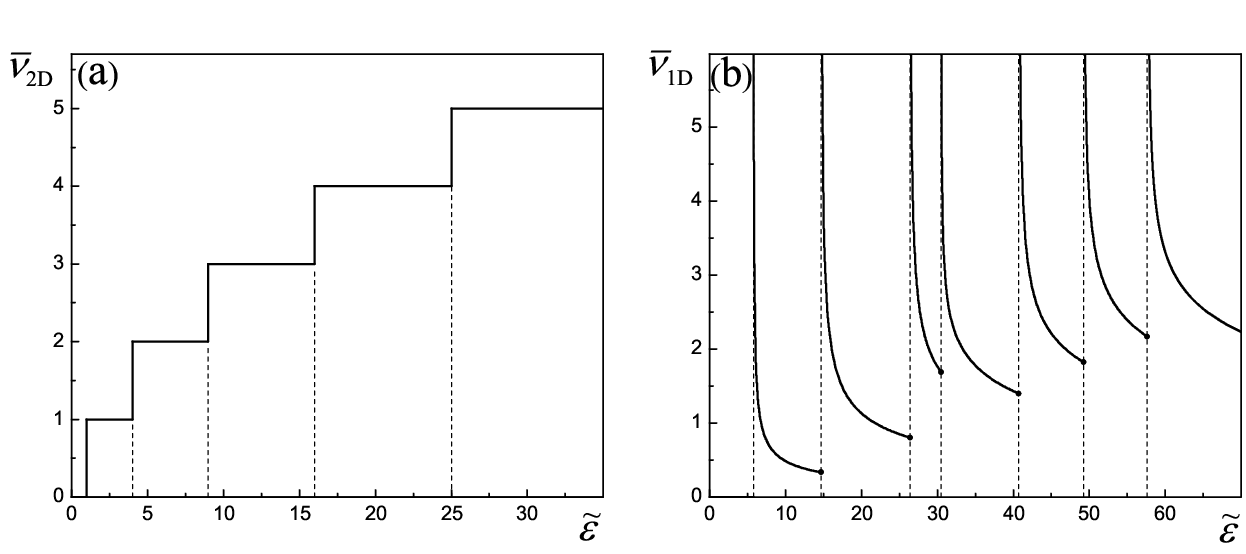} 
\vspace{-6mm}
\caption{\label{fig07} %
The density of states for the quasi-two-dimensional quantum well and the nanotube: \newline %
(a) $\bar{\nu}_{\rm{2D}}(\tilde{\varepsilon})=\nu_{\rm{2D}}(\tilde{\varepsilon})\left(\frac{mA}{\pi\hbar^2}\right)^{\!-1}\!=\sum_{n=1}^\infty\theta(\tilde{\varepsilon}\!-\!n^2),  %
\,\tilde{\varepsilon}=\varepsilon\big/\varepsilon_L,\, \varepsilon_L\!\equiv\!\frac{\pi^2\hbar^2}{2mL_0^2}$, %
$A$ is the area, $L_0$ is the distance between planes \cite{PS}; \newline %
(b) $\bar{\nu}_{\rm{1D}}(\tilde{\varepsilon})=\nu_{\rm{1D}}(\tilde{\varepsilon})\left(\frac{2mLR}{\pi\hbar^2}\right)^{\!-1}\!=Z_{-1/2}^{(r_0)}(\tilde{\varepsilon})$  %
\,at\, $\rho_{r_0}^2<\tilde{\varepsilon}\leq\rho_{r_0+1}^2$, $\tilde{\varepsilon}=\varepsilon\big/\varepsilon_R$. %
The dashed lines correspond to $\tilde{\varepsilon}=\rho_r^2$.
}%
\end{figure*}

\section{Dependencies of thermodynamic quantities on the tube radius} %
If thermodynamic quantities are taken in the reduced form (\ref{19}),
and the dimensionless chemical potential $\eta$ and the dimensionless temperature $\tau$
are used as independent variables, then as it was shown
the geometrical dimensions fall out of the thermodynamic relations,
in particular the radius of the tube $R$ falls out.
Meanwhile, exactly the dependencies of the observable quantities on the tube radius
are of interest in experiment. To obtain such dependencies, the relations derived above
should be presented in the dimensional form.
At that, the form of dependence of the thermodynamic quantities on the radius
will essentially depend on what quantity is being fixed when studying such dependence:
the total density $n$ or the linear density $n_{\!L}$.
Let us show it on the example of dependence of the chemical potential on the tube radius.
In the dimensional form the formulas for the densities can be presented as follows
\begin{equation} \label{53}
\begin{array}{ll}
\displaystyle{%
  \pi^{3/2}R^3n=\pi^{1/2}R\,n_{\!L}=\frac{2}{\sqrt{\pi}}\,Z_{1/2}^{(r_0)}(\eta), %
}%
\end{array}
\end{equation}
and, according the definition, the chemical potential is expressed through the parameter $\eta$ by the formula
\begin{equation} \label{54}
\begin{array}{ll}
\displaystyle{%
  \mu=\frac{\hbar^2}{2mR^2}\,\eta. %
}%
\end{array}
\end{equation}
The formulas (\ref{53}) and (\ref{54}) define parametrically
the dependence of the chemical potential on the radius
in the domain of variation of the parameter $\rho_{r_0}^2<\eta<\rho_{r_0+1}^2$.
As we see, the dependence of $R$ on the parameter $\eta$ will be different
depending on what is fixed -- the volume or the linear density.

At the fixed linear density $n_{\!L}=\rm{const}$ in the domain $\rho_{r_0}^2<\eta<\rho_{r_0+1}^2$
the dependence of the chemical potential on the radius is parametrically defined by the equations
\begin{equation} \label{55}
\begin{array}{ll}\hspace{0mm}
\displaystyle{%
  \bar{\mu}=\frac{\eta}{\big[Z_{1/2}^{(r_0)}(\eta)\big]^2}, \quad %
  \bar{R}=Z_{1/2}^{(r_0)}(\eta),
}
\end{array}
\end{equation}
where
\begin{equation} \label{56}
\begin{array}{ll}\hspace{0mm}
\displaystyle{%
  \bar{R}=\frac{\pi n_{\!L}}{2}R, \quad %
  \bar{\mu}=\frac{2m}{\hbar^2}\Big(\frac{2}{\pi n_{\!L}}\Big)^{\!2}\mu. %
}
\end{array}
\end{equation}
At the fixed total density $n=\rm{const}$ the similar dependence is defined by the equations
\begin{equation} \label{57}
\begin{array}{ll}\hspace{0mm}
\displaystyle{%
  \bar{\mu}=\frac{\eta}{\big[Z_{1/2}^{(r_0)}(\eta)\big]^{2/3}}, \quad %
  \bar{R}=\big[Z_{1/2}^{(r_0)}(\eta)\big]^{1/3},
}
\end{array}
\end{equation}
where
\begin{equation} \label{58}
\begin{array}{ll}\hspace{0mm}
\displaystyle{%
  \bar{R}=\Big(\frac{\pi^2 n}{2}\Big)^{\!1/3}R, \quad %
  \bar{\mu}=\frac{2m}{\hbar^2}\Big(\frac{2}{\pi^2 n}\Big)^{\!2/3}\mu. %
}
\end{array}
\end{equation}
The dependencies of the chemical potential on the tube radius
are shown for both cases in Fig.\,8. As seen, at the fixed linear density
the chemical potential monotonically decreases with increasing the radius (curve {\it 1} in Fig.\,8).
The oscillation dependence on the radius is realized at the fixed total density (curve {\it 2} in Fig.\,8).
The amplitude of oscillations rapidly decreases with increasing the radius,
at that the chemical potential quite slowly approaching its volume value
$\displaystyle{\bar{\mu}_{\rm{3D}}=\lim_{\bar{R}\rightarrow\infty}\bar{\mu}=6^{2/3}}$.

Let us also give the dependencies on the radius of such a directly measurable quantity as the pressure.
At the fixed linear density $n_{\!L}=\rm{const}$ in the domain $\rho_{r_0}^2<\eta<\rho_{r_0+1}^2$ %
the dependencies of the radial and the parallel pressures on the parameter $\eta$ are defined by the formulas:
\begin{equation} \label{59}
\begin{array}{ll}
\displaystyle{\hspace{-1mm}%
\hspace{0mm}
  \bar{p}_R=\frac{\eta Z_{1/2}^{(r_0)}(\eta)-Z_{3/2}^{(r_0)}(\eta)}{\big[Z_{1/2}^{(r_0)}(\eta)\big]^5},\quad %
  \bar{p}_\parallel=\frac{2}{3}\frac{Z_{3/2}^{(r_0)}(\eta)}{\big[Z_{1/2}^{(r_0)}(\eta)\big]^5},%
}%
\end{array}
\end{equation}
where
\begin{equation} \label{60}
\begin{array}{ll}
\displaystyle{%
  \bar{p}_{R,\parallel}=\pi^2\frac{m}{\hbar^2}\Big(\frac{2}{\pi n_{\!L}}\Big)^{\!5}p_{R,\parallel} %
}%
\end{array}
\end{equation}
and the dependence of the radius on $\eta$ is given by the formulas (\ref{55}),(\ref{56}).
At the fixed total density $n=\rm{const}$ the similar dependencies are defined by the equations
\begin{equation} \label{61}
\begin{array}{ll}
\displaystyle{\hspace{-0mm}%
\hspace{0mm}
  \bar{p}_R=\!\frac{\eta Z_{1/2}^{(r_0)}(\eta)-Z_{3/2}^{(r_0)}(\eta)}{\big[Z_{1/2}^{(r_0)}(\eta)\big]^{5/3}},\,\,\, %
  \bar{p}_\parallel=\!\frac{2}{3}\frac{Z_{3/2}^{(r_0)}(\eta)}{\big[Z_{1/2}^{(r_0)}(\eta)\big]^{5/3}},%
}%
\end{array}
\end{equation}
where
\begin{equation} \label{62}
\begin{array}{ll}
\displaystyle{%
  \bar{p}_{R,\parallel}=\pi^2\frac{m}{\hbar^2}\Big(\frac{2}{\pi^2 n}\Big)^{5/3}p_{R,\parallel}\,, %
}%
\end{array}
\end{equation}
and the dependence of the radius on $\eta$ is given by the formulas (\ref{57}),(\ref{58}).
The dependencies of the pressures on the tube radius are shown for both cases in Fig.\,9.
\vspace{0mm} %
\begin{figure}[t!]
\vspace{-6mm} \hspace{-0mm}
\includegraphics[width = 1.0\columnwidth]{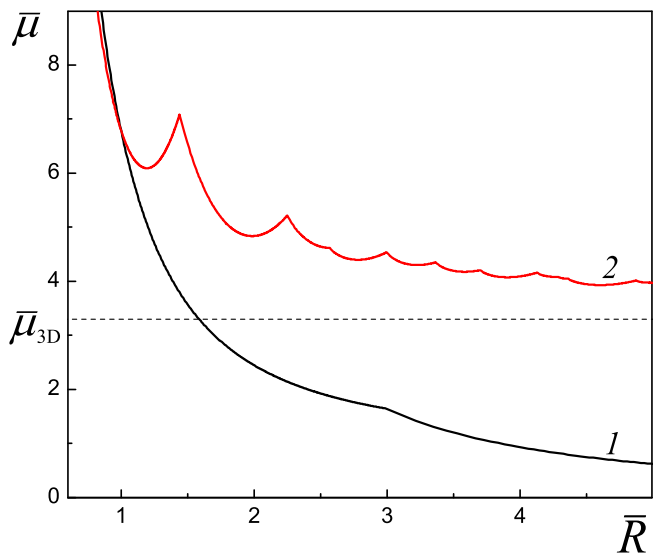} %
\vspace{-9mm} 
\caption{\label{fig08} %
The dependencies of the chemical potential on the tube radius:
\!({\it 1}) at the fixed linear density $n_{\!L}$ ($\bar{\mu}, \bar{R}$ are defined by formulas (\ref{56})); %
\!({\it 2}) at the fixed volume density $n$ ($\bar{\mu}, \bar{R}$ are defined by formulas (\ref{58})). %
The dashed line shows the chemical potential
$\displaystyle{\bar{\mu}_{\rm{3D}}\!=\!\lim_{\bar{R}\rightarrow\infty}\bar{\mu}=6^{2/3}}$ of volume system of density $n$. %
}%
\end{figure}

Here also, the oscillations exist only under condition of the fixed total density,
so that qualitatively the situation in this quasi-one-dimensional case is similar to that which
takes place in the quasi-two-dimensional quantum well \cite{PS}. 
At the fixed linear density $n_{\!L}$ with increasing the radius the total density $n$ approaches zero,
so that the both pressures approach zero as well.
In the opposite case $R\rightarrow 0$ the total density increases unlimitedly
and therefore the pressures also tend to infinity (curves {\it 1},\,{\it 2} in Fig.\,9).
At the fixed total density $n$ with increasing the radius the both pressures
approach, oscillating, their volume value
$\displaystyle{\bar{p}_{\rm{3D}}=\!\lim_{\bar{R}\rightarrow\infty}\bar{p}_{R,\parallel}=(2/5)6^{2/3}}$. %
In the limit $R\rightarrow 0$\, it should be $L\rightarrow\infty$, and so
the linear density $n_{\!L}\rightarrow 0$.
Hence the parallel pressure also approaches zero $p_\parallel\rightarrow 0$. %
The radial pressure with decreasing the tube radius tends to infinity $p_R\rightarrow\infty$
due to the zero-point oscillations in the radial direction (curves {\it 3},\,{\it 4} in Fig.\,9).

\vspace{0mm} %
\begin{figure}[t!]
\vspace{-6mm} \hspace{-5mm}\vspace{0mm}
\includegraphics[width = 1.03\columnwidth]{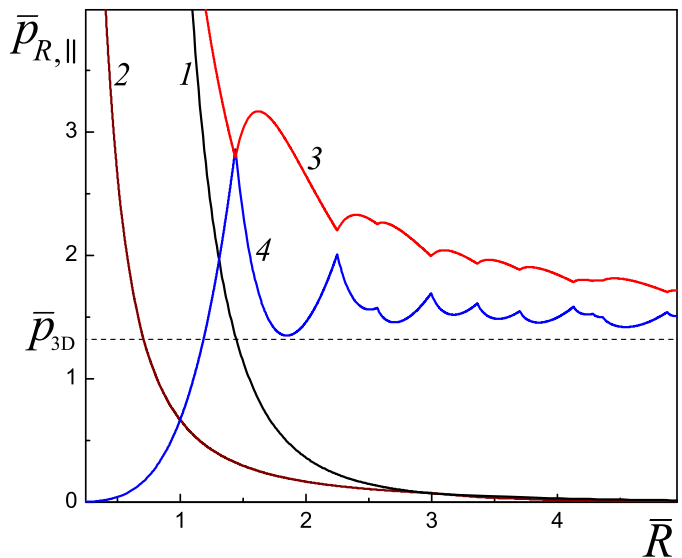} %
\vspace{-5mm} 
\caption{\label{fig09} %
The dependencies of the pressures on the tube radius: %
\!({\it 1}) $\bar{p}_R(\bar{R})$ and ({\it 2}) $\bar{p}_\parallel(\bar{R})$ at the fixed linear density $n_{\!L}$; %
({\it 3}) $\bar{p}_R(\bar{R})$ and ({\it 4}) $\bar{p}_\parallel(\bar{R})$ at the fixed volume density $n$. %
The dashed line shows the pressure
$\displaystyle{\bar{p}_{\rm{3D}}=\!\lim_{\bar{R}\rightarrow\infty}\bar{p}_{R,\parallel}=(2/5)6^{2/3}}$
of volume system of density $n$. %
}%
\end{figure}

\vspace{-0.5mm}
\section{Conclusion}\vspace{-1.0mm}
The exact formulas for calculation of the thermodynamic functions of
the ideal Fermi gas that fills the tube of an arbitrary radius at arbitrary temperature
have been derived in the work.
It is shown that all thermodynamic quantities,
written in the dimensionless reduced form not containing geometric dimensions,
can be expressed through some standard functions (\ref{06})
of the dimensionless temperature and chemical potential and their derivatives. %
The thermodynamic potential, energy, density, entropy, equations of state,
heat capacities and compressibilities of the Fermi gas at arbitrary temperatures
in the considered quasi-one-dimensional case are calculated through the introduced standard functions.
It is shown that owing to the anisotropy of the system the Fermi gas has two equations of state
since the pressures along the tube and in the direction of its radius are different.
Due to the same reason the system is characterized by more number of heat capacities than in the volume case.

At low temperatures the entropy and all heat capacities are linear in temperature everywhere except the specific points
where the chemical potential coincides with the square of the Bessel function zero and the filling of a new discrete level begins.
At these specific points the root dependence of the entropy and heat capacities on temperature appears.
The dependencies of the entropy and heat capacities on the chemical potential have sharp maximums at the specific points.
This differs the quasi-one-dimensional case from the the quasi-two-dimensional where the heat capacities and entropy undergo jumps
at the beginning of the filling of a new discrete level.
The dependencies of the pressures and compressibilities on the chemical potential at zero temperature are calculated.
It is shown that the character of dependence of thermodynamic quantities on the tube radius
essentially depends on whether this dependence is considered at the fixed linear or the fixed total density.
At the fixed linear density the thermodynamic quantities vary monotonically with the radius,
and at the fixed total density they undergo oscillations.\!
We believe that the results of this work and work \cite{PS} devoted to the quasi-two-dimensional system
can serve as a basis for further consistent study of low-dimensional systems of interacting Fermi particles.

\end{document}